\documentstyle[12pt,fleqn]{article}
\font\fontb=cmr12 scaled\magstep2 % 1.2*1.2
\font\fontc=cmr12 scaled\magstep1 % 1.2
\def\beq {\begin{eqnarray}}
\def\eeq {\end{eqnarray}}
\def\beqn {\begin{eqnarray*}}
\def\eeqn {\end{eqnarray*}}
\def\neqn {\nonumber}
\def\bs {\indent\indent}
\def\in {\indent}
\def\ni {\noindent}
\def\new {\newpage}
\def\PL #1 #2 #3 {Phys. Lett.~{\bf#1} (#2) #3}
\def\NP #1 #2 #3 {Nucl. Phys.~{\bf#1} (#2) #3}
\def\ZP #1 #2 #3 {Z.~Phys.~{\bf#1} (#2) #3}
\def\PR #1 #2 #3 {Phys. Rev.~{\bf#1} (#2) #3}
\def\PP #1 #2 #3 {Phys. Rep.~{\bf#1} (#2) #3}
\def\PRL #1 #2 #3 {Phys. Rev.~Lett.~{\bf#1} (#2) #3}
\def\PTP #1 #2 #3 {Prog. Theor.~Phys.~{\bf#1} (#2) #3}
\def\MPL #1 #2 #3 {Mod. Phys.~Lett.~{\bf#1} (#2) #3}
\def\IJM #1 #2 #3 {Int. J.~Mod.~Phys.~{\bf#1} (#2) #3}
\def\etal {{\it et al}.}
\def\eg {{\it e.g}.}
\def\ie {{\it i.e}.}

\def\qcon{$ \langle {\bar \psi }\psi \rangle $}

\def\q4{${\bar \psi} \Gamma_i \psi{\bar \psi} \Gamma_j \psi$}

\def\bra {\langle}
\def\ket {\rangle}
\def\GeV{\mbox{GeV}}
\def\MeV{\mbox{MeV}}
\def\fm{\mbox{fm}}

\def\tr {\mbox{tr}}

\def\q   {{\bf q}}

\begin{document}
%%%%%%% TITLE PAGE %%%%%%%%%%%%%%
\begin{titlepage}
%\vspace* {0.1cm}
%
\begin{flushright}{TMU-NT950501, hep-ph/9505325
\\May 1995}
\end{flushright}
%}
\vspace {0.2cm}
\baselineskip=1cm
\begin{center}{\fontb Pion structure function in nuclear medium} \\
\vspace {3.1cm}
{\fontc Katsuhiko Suzuki\footnote{e-mail address :
ksuzuki@atlas.phys.metro-u.ac.jp}}\\
%address after September 1, 1995: Department of Physics,
%Technical University of Munich,
%D-85747 Garching, Germany}}\\
{\em Department of Physics, Tokyo Metropolitan University}\\
{\em Hachiohji, Tokyo 192-03, Japan}

\vspace {2.2cm}
{\bf Abstract}
\end{center}
\vspace {0.5cm}

\baselineskip=0.8cm
\noindent
We study the pion structure function in nuclear medium using the
Nambu and Jona-Lasinio model, and its
implication for the nuclear pion enhancement of the sea quark
distribution in nuclei.
By using the operator product expansion, medium effect of the nuclear
matter is incorporated in calculations of the twist-2 operators.
We find density dependence of the pion structure function is
rather weak around the nuclear matter density.
We also discuss how the medium modification of the pion structure
affects the sea quark enhancement in the nucleus.

\vspace{0.3cm}
%PACS numbers: 24.85.+p, 12.39.Fe, 13.60.Hb, 12.39.Ki

\end{titlepage}

%-------------%-------------%-------------%-------------%------------
\newpage
\baselineskip = 0.75cm

\section{Introduction}
\bs
Modification of the hadron structure in nuclear medium is
an interesting topic of the nuclear physics, and has attracted
considerable interests.    In the deep inelastic scattering, such
modifications of the
nucleon structure in nuclei are observed experimentally, that is, EMC
effect\cite{EMC}.
It is very difficult to calculate the hadron structure function
itself by using
QCD directly, though the measured momentum dependence of the structure
function is
consistent with predictions of the perturbative QCD.   Recently,
several attempts were done to calculate the structure function in
terms of
low energy effective theories of QCD, which reproduce hadron
properties such as masses and magnetic moments\cite{Jaffe-Ross}.
In those studies, it is assumed
that calculation of the twist-2 matrix elements within the effective
theory gives  structure function at certain low energy scale
$\mu \sim 1 \GeV$,
where the effective theories make sense.
The resulting structure function is evoluted to the experimental high
momentum scale with the help of the perturbative QCD,
and comparison with experiments can be made\cite{Models1,Models2}.

In particular, the pion structure function
is of special interest so as to study the nuclear pion
enhancement of the sea quark distribution in the deep
inelastic scattering\cite{Smith,ET} and the Drell-Yan (DY)
process\cite{pion_DY}.
The nuclear pion enhancement due to the attractive nucleon-nucleon
interaction in the nucleus was extensively studied in the
last decade\cite{OTW}.
The nucleon-hole and delta-hole excitations shown in Fig.1 modify
the pion-like mode in the nucleus,
and lead an excess of the pion number in nuclei.
Llewellyn Smith\cite{Smith}, and Ericson and Thomas\cite{ET}
discussed that such an enhancement of the pion number changes the
sea quark distribution in the nucleus.
Assuming that a part of the nucleon sea quark distribution is
originated
from the virtual pion cloud, the sea quark distribution of the
nucleus is written by the convolution formula\cite{Sullivan};
\beq
\bar q_{\pi NN}^{A}(x,Q^2)=\int_x^1 {dy / y}\, f_{\pi NN}^{A}(y)\,
\bar q_\pi ^{}(x / y,Q^2) \;\; ,
\label{convl}
\eeq
where $\bar  q_{\pi NN} ^A (x,Q^2)$ is the sea quark distribution of
the nucleon in the nucleus
originated from the pion cloud, and  $\bar {q}_\pi(x_\pi,Q^2)$ the
antiquark distribution in the pion.
Following the work of Ericson and Thomas, the longitudinal pion
momentum distribution in the nucleus $f_{\pi NN}^{A}(y)$ is
given by\cite{ET},
\beq
f_{\pi NN}^A (y)=3 \; y {{g_{\pi NN}^2}
\over {16\pi ^2}}\int_{m_N^2 y}^\infty  {dq}
\int_0^{q -m_N y} d \omega \;
{q^2 \over {(t+m_\pi ^2)^2}}[F_{\pi NN}^{}(q^2)]^2  \,
R(\omega, q) \;\;,
\label{pion_nucl}
\eeq
where $g_{\pi NN}$ is the pion nucleon coupling constant, $F_{\pi
NN}^{}(t)$ the pion-nucleon formfactor, and $m_N$, $m_\pi$ the
masses of the nucleon and pion, respectively.
Here, $R(\omega, q)$ is the response function of the nuclear matter
calculated by the random phase approximation,
and carries the information
of the collective excitation illustrated in Fig.1\cite{OTW}.
In the case of free nucleon, we basically replace $R(\omega, q)$
with 1,
and recover the Sullivan's convolution formula for the single nucleon.

By using (\ref{convl}), substantial enhancement of the sea quark
distribution in
nuclei was found by several authors\cite{ET,pion_excess},
and such an enhancement should be
observed in the DY experiment\cite{pion_DY}.
However, recent experiment of the DY process indicates that there
is no
enhancement of the sea quark in nuclei\cite{DY_Exp}.
Also, the sea quark enhancement due
to the pion excess is not observed in the deep inelastic
scattering\cite{DIS_Exp}.
Recently, Bertsch {\etal}\cite{NPE_Bertsch} and Brown
{\etal}\cite{NPE_Brown} studied such a puzzle `why the
nuclear pion enhancement is not observed', by taking into account the
modification of the gluon properties in the nucleus, or the universal
scaling law due to the chiral symmetry restoration in the
medium\cite{BR}.

In the above discussion, the medium effect leads an excess of the
number of pion in nuclei, but the pion
structure function itself is assumed to be unchanged in the
nuclear matter.
In this letter, we concentrate on the medium effect on {\em the quark
distribution
in the pion}, and its implication for the nuclear pion enhancement.
If the medium effects change the pion distribution in the nucleus,
it is
also important to examine whether the nuclear medium modifies
the quark distribution in the pion or not.
Following this scenario, we study the density dependence of the
pion structure function.

The antiquark distribution of the pion appeared in (\ref{convl})
consists of the valence antiquark and sea quarks.  For
$x>0.1$\footnote{ We do not consider the region $x< 0.1$, since
the nuclear shadowing effect, which we do not take into account
in this
letter, is important in that region.}, most contribution
to the convolution formula comes from the valence
antiquark part.  Hence, we consider only the valence quark distribution
in the pion and its medium modifications in this letter.

In the previous studies of the present author\cite{Shige1,Shige2},
we have evaluated the pion
structure function using the Nambu and Jona-Lasinio model (NJL)
\cite{NJL}, and obtained
reasonable agreements with experimental data.
Here, we adopt the similar procedure in the
nuclear medium.  The NJL model, where the chiral invariance is main
ingredient, describes the dynamical mass generation of the constituent
quark, and reproduces $SU(3)$ meson properties successfully
as a low energy effective theory of QCD\cite{Review_NJL}.
It is believed that restoration of the chiral symmetry takes place
at high density.  Even at the nuclear matter
density $\rho=\rho_0 = 0.17\fm ^{-3}$, the chiral symmetry may
be partially restored.
We will show how the pion structure function changes in the nuclear
medium in terms of the NJL model.

%----------%----------%----------%----------%----------
\vspace{0.3cm}
\section{Nambu and Jona-Lasinio model in medium}
\bs
Before we start calculating the pion structure function in medium, we
briefly give a basic procedure of the NJL model
at finite density.  The NJL model demonstrates the spontaneous chiral
symmetry breaking of the QCD vacuum, in which the gluon degrees of
freedom are assumed to be frozen into a chiral invariant 4-quark
interaction.
The NJL lagrangian is written as,
\beq
{\cal L}_{NJL}&=&\bar{\psi}(i\gamma^{\mu}\partial_{\mu}-m)\psi
          +G_{S}[(\bar{\psi}t_{i}\psi)^2
                  +(\bar{\psi}i\gamma_{5}t_{i}\psi)^2] \;\; ,
\label{NJL_lag}
\eeq
where \(\psi\) denotes the quark field with current mass \(m\),
$t_{i}$ are $SU(3)$ flavor matrices with the normalization
$\tr [t_i t_j] = \delta _{ij} /2$, and \(G_S\) the coupling constant.
In this model, the constituent quark mass $M$ and the quark condensate
{\qcon} have non-zero values due to the dynamical chiral symmetry
breaking, if the coupling constant $G_S$ is larger than a certain
critical value.
As a result, the zero mass Goldstone pion appears in the chiral
limit $m=0$\cite{NJL}.  Using the realistic current quark masses, $m_u
\sim 5\MeV$ and $m_s \sim 150 \MeV$,  $SU(3)$ meson
properties are well reproduced in the generalized NJL model with the
$U(1)_A$ anomaly\cite{Review_NJL}.

For the application at finite density, we adopt the quark propagator;
\beq
S(p)=(\not p+M)\left[ {{1 \over {p^2-M^2+i\varepsilon }}
+{{i\pi } \over E}\delta (p^0-E)\,\theta (k_F- |\vec p |)} \right]
\label{Dpropahator}
\eeq
where $E= \sqrt{p^2 + M^2}$.
The quark Fermi momentum $k_F$ is related with the baryon number
density.  Restricting the 2-flavor, the baryon number density $\rho$
is given by,
\beqn
\rho = 4 \int^{k_F} _0
{\frac{d^3k}{(2\pi)^3}} \;.
\eeqn

The quark self-energy is modified by the presence of other
quarks in medium.  From (\ref{Dpropahator}), the quark condensate
becomes,
\beq
\langle \bar \psi \psi  \rangle &=&-i \tr \int {{{d^4p}
\over {(2\pi )^4}}}S(p)\neqn\\
&=& - \frac{N_c}{\pi^2} \int^\Lambda _{k_{F}} {dp \,p^2} \frac{M}
{\sqrt{p^2 + M^2}} \;.
\label{Dcondensate}
\eeq
Then, the Gap equation depends on the Fermi momentum.
\beq
M = m - G_S  \langle  \bar{\psi} \psi  \rangle
\label{Gap}
\eeq
Apparently, the constituent quark mass $M$ vanishes in the limit
$k_F \to \infty$.
In this letter, we use the 3-dimensional sharp cutoff scheme.
Choice of the 3-dim. sharp cutoff makes it possible to maintain
the gauge
invariance in the following calculations\cite{Review_NJL}.

The pion properties are also obtained by solving the Bethe-Salpeter
(BS) equation.  The quark-antiquark scattering matrix in the pion
channel with the total momentum $p$ is evaluated as,
\beq
{\cal T}(p) = \frac {2 G_S} {1 - 2 G_S \Pi(p^2) }
\label{T-pion}
\eeq
where
\beq
\Pi(p^2) = i \tr \int {{{d^4 k}
\over {(2\pi )^4}}} [i \gamma_5 S(k) i \gamma_5 S(k+p)] \neqn
\eeq
The pion mass is determined by the pole of denominator at
$p^2 = m_\pi ^2$;
\beqn
1 - 2 G_S \Pi(p^2 = m_\pi ^2) = 0
\eeqn
Straightforward calculation yields,
\beq
\frac{m} {M} + 2 G_S \, p^2 I_2 (p^2) = 0 \;\; ,
\eeq
where
\beq
I_2(p^2 ) = -\frac{N_c}{4 \pi^2} \int ^{4( M^2 + \Lambda^2)}
_{4 (M^2 + {k_F}^2)} \frac{1}{ \kappa^2 - p^2} \sqrt{1 - \frac{4
M^2}{\kappa^2}}  \;\; .
\eeq

Form  the scattering matrix (\ref{T-pion}), the pion-quark-quark
coupling constant is expressed as,
\beq
g^2_{\pi qq} = \frac {1}{- \frac{d}{dp^2}[ p^2 I_2 (p^2)]}  \;\; .
\label{gpqq}
\eeq
We use the parameter set, $m=5.8\MeV$,
$\Lambda=615\MeV$, $G_S \Lambda^2 = 9.08$, which are chosen to reproduce
the pion mass $140\MeV$ and the decay  constant $f_\pi = 93.3 \MeV$ at zero
density.
Results for the constituent quark mass and the quark condensate are
$M=358 \MeV$ and $\bra \bar \psi \psi \ket ^{1/3} = -245 \MeV$,
which are consistent with phenomenological values.
All the results are basically same as
those obtained in previous studies\cite{Review_NJL}.

In the nuclear medium, the quark condensate decreases as the baryon
number density increases, and hence the
constituent mass and the pion decay constant also decrease.
At the normal nuclear matter density, $\rho_0 =0.17 \fm^{-3}$,
values of these
quantities become about $70 \%$ of their zero density values.
The pion mass is almost independent of the density before the
chiral phase transition.  After the chiral symmetry is restored,
the pion mass grows rapidly.
Such a behavior is general tendency of the NJL model
calculations\cite{Kle}.

%----------%----------%----------%----------%----------
%\vspace{0.5cm}
\section{Calculation of the moments of the structure function}
\bs
We shall consider the medium modification of the pion structure function
using the NJL model.
In order to calculate the pion structure function,
we adopt a formalism based on the operator product expansion (OPE)
to separate the hard and soft parts of the matrix elements
explicitly\cite{Muta}.
By doing so, it is easy to deal with the effects of finite density
on the matrix elements.

As a result of the OPE, the forward scattering amplitude is written
by the hard part
(Wilson coefficients) and the matrix elements of local operators as
illustrated in Fig.2.
The twist-2 operators of the un-polarized
structure function are defined as,
\beq
O_{{\mu_1}{\mu_2} \cdots {\mu_n}} = \frac{i^n}{n!} \bar \psi
\gamma_{\mu_1} D_{\mu_2} \cdots D_{\mu_n} \psi + \mbox{ permutations}
\eeq
where $D_\mu$ is the covariant derivative of QCD.  Here, we write the
quark part only.   Also define the
matrix element $A_n$,
\beq
\bra p |
O_{{\mu_1}{\mu_2} \cdots {\mu_n}} | p \ket =
A_n p_{\mu_1}  p_{\mu_2} \cdots  p_{\mu_n} + \mbox{trace terms}
\eeq
where $ | p \ket$ is the pion state with the momentum $p$.  We also
write the momentum of the virtual photon $q$, $\nu = p \cdot q$, and
the Bjorken-$x$; $x = -q^2 /2 m_\pi \nu$.
This matrix element is related with the moment of the
structure function;
\beq
\int^1_0 {dx} \; x^{n-2} F_2 (x,q^2) = A_n (\mbox{log}[q^2])
\eeq
where the structure function $F_2 (x) = \sum {e^2_i} x q_i (x)$ with
the $q_i (x)$ being the quark longitudinal momentum distribution
function in the pion.

Here, we choose the axial gauge $q \cdot A_\mu= 0$.  Hence,
multiplication of $q_{\mu_1} \cdots q_{\mu_n}$ for both sides yields,
\beq
A_n \nu^n = \bra p | \bar \psi \not q (q\cdot \partial )^{n-1}
 \psi | p \ket
\eeq
Then, we can get expressions for the $n$-th moment of the structure
function.

We shall evaluate the matrix element of local operators in terms
of the NJL model.
The diagram Fig.2 indicates only the valence quark contribution
up to the leading order.
Hereafter, we stress on the $n$-th moment of the valence quark
$A_n^{val}$, which is related with the valence quark
distribution as,
\beq
A_n^{val} = \int^1_0 {dx} \; x^{n-1} q(x) .
\label{Mellin}
\eeq
Throughout the following calculations,
we omit the overall charge (isospin) factor for simplicity.
The valence antiquark also has a similar contribution.

$A_1^{val}$ is written by,
\beq
\hspace{-0.3cm} A_1^{val} p_\mu = g_{\pi qq}^2 \tr \int {{{d^4k} \over {(2\pi
)^4}}}
{{[i\gamma _5(\not k+\not p+M)\gamma _\mu (\not k+\not p+M)
i\gamma _5(\not k+M)]} \over {(k^2-M^2+i\varepsilon )[(k+p)^2-M^2+i
\varepsilon ]^2}}
\eeq
After some algebra, we find
\beq
A_1^{val} p_\mu &=& -g_{\pi qq}^2 p_\mu
[p^2 \frac{d}{d p^2} I_2(p^2) + I_2(p^2)] \neqn \\
&=&  - p_\mu g_{\pi qq}^2 \frac{d}{d p^2 } [p ^2 I_2(p^2)]  \neqn \\
&=& p_\mu \;.
\label{1st_momA}
\eeq
where we have used eq. (\ref{gpqq}) for $g_{\pi qq}$ in the last
equality.  Thus, we get $A_1^{val}=1$, which manifests the
Adler sum rule for the valence quark distribution function.
Antiquark contribution also gives the same result.

We next consider the second moment which gives the momentum fraction
carried by quarks.   The 2-nd moment is defined by,
\beq
A_2^{val} p_\mu p_\nu = \frac{1}{2!} [ \bra p | \bar \psi \gamma_\mu
 (k+p)_\nu  \psi | p \ket + \mbox{permutation}] \;,
\label{2nd_mom}
\eeq
Similar manipulation with the case of the 1-st moment gives,
\beqn
&\sim &\int_{}^{} {{{d^4k} \over {(2\pi )^4}}} [
{{{k_\mu k_\nu +k_\mu p_\nu } \over {[(k+p)^2-M^2+i\varepsilon ]^2}}}
{+{{p_\mu (k+p)_\nu } \over {[(k+p)^2-M^2+i\varepsilon ]^2}}} \\
 &&\hspace{1cm} {+p_\mu {d \over {dp^2}}{{p^2k_\nu } \over
{(k^2-M^2+i\varepsilon )((k+p)^2-M^2+i\varepsilon) }}}  \\
&&\hspace{1.5cm} {+p_\mu p_\nu {d \over {dp^2}}{{p^2}
\over {(k^2-M^2+i\varepsilon )
((k+p)^2-M^2+i\varepsilon )}}} ]
\eeqn
The first and second term give the trace terms, which are proportional
to $g_{\mu\nu}$, and are irreverent in the Bjorken limit\cite{Muta}.
Sum of the third and forth terms becomes,
\beq
A_2^{val} p_\mu p_\nu &=& - p_\mu p_\nu g_{\pi qq}^2
\frac{d}{d p^2} p^2 I_2
\neqn \\
&& \hspace{0.25cm} - p_\mu g_{\pi qq}^2  \frac{d}{d p^2} [ p^2
\int {{d^4k} \over {(2\pi )^4}}
\frac {k_\nu} {(k^2-M^2+i\varepsilon )[(k+p)^2-M^2+i\varepsilon ]} ]
\neqn \\
& = & - p_\mu p_\nu g_{\pi qq}^2 \frac{d}{d p^2}
[ p^2 (I_2 -  \frac{1}{2} I_2)] \neqn \\
& = & \frac{1}{2} p_\mu p_\nu \;.
\label{2nd_momA}
\eeq
Again, the calculation for the antiquark part gives 1/2.  This is just
the momentum sum rule.
\beqn
\int^1_0{dx} x [q(x) + \bar q (x)] = 1
\eeqn
Above result shows that the momentum sum rule is strictly satisfied
in the NJL
model, {\ie} all the momentum in the pion is carried by the valence
quarks at the low energy model scale.  Of course, if we consider sea
quark
degrees of freedom, which are identified with higher order loop
corrections in the
NJL model\cite{Shige2,MW}, the sea quarks carry the certain amount of
the momentum fraction even at the model scale.

In the previous studies\cite{Shige1,Shige2}, we had adopted the
covariant
parton formalism and artificial cutoff procedure, which breaks
the gauge invariance.
The correct treatment described here maintains the momentum sum rule.

The results for the 1-st and 2-nd moments are quite reasonable.
Characteristic feature of the non-perturbative model may be seen in
higher moments.   It is worth mentioning that the 1-st and 2-nd
moments are independent of the density.

Calculations for the higher moments require more tedious manipulations.
The 3-rd moment is defined by,
\beq
A_3^{val} p_\mu p_\nu p_\gamma = \frac{1}{3!} [ \bra p | \bar
\psi \gamma_\mu
 (k+p)_\nu (k+p)_\gamma \psi | p \ket + \mbox{permutation}] \;.
\label{3rd_mom}
\eeq
We show only the result.
\beq
A_3^{val} = \frac{1}{4} + g^2_{\pi qq} {{N_c} \over {8\pi ^2}}
I_3 (p^2,M)
\label{3rd_momA}
\eeq
\ni
where
\beq
I_3 (p^2,M)&=&\int\limits_{4(M^2+k^2_F )}^{4(M^2+\Lambda ^2)}
 {d\kappa ^2}\sqrt {{{\kappa ^2-4M^2}
 \over {p^2}}}\neqn \\
&&\hspace{1cm}
 \times [{-{1 \over {8p^2}}\log \left| {{{\sqrt {p^2}+
 \sqrt {\kappa ^2}}
 \over {-\sqrt {p^2}+\sqrt {\kappa ^2}}}} \right|}
  + {1 \over 4}{1 \over {\kappa ^2-p^2}}
\sqrt {{{\kappa ^2} \over {p^2}}} ]
\label{I3}
\eeq
\ni
Note that the 3-rd moment explicitly depends on the NJL model
parameters. Whenever the density increases, the 1-st and 2-nd
moments are unchanged, {\ie} they are just numerical numbers.
However, the 3-rd moment shows density
dependence through the constituent quark mass, coupling constant,
and pion
mass.  The restoration of the chiral symmetry drives the
reduction of the 3-rd moment, which will be shown later.

Similarly, the 4-th moment is calculated as,
\beq
A_4^{val} = \frac{1}{8} + g^2_{\pi qq}
{{3N_c} \over {16 \pi ^2}} I_3 (p^2,M) \;\; .
\label{4th_momA}
\eeq
Higher moments are calculable in the same manner.

It is interesting to note that numerical numbers appeared in eqs.
(\ref{3rd_momA}) and (\ref{4th_momA}), 1/4 and 1/8, are the same as
results of the delta function quark distribution $q(x) =
\delta(x-1/2)$.  Such expressions are desirable.
Remaining model dependent parts
are corrections due to the non-perturbative bound state structure.
In the limit of high density $\rho \to \infty$,
the correction terms go to zero, and the
structure function approaches the non-relativistic (non-interacting)
 limit $q(x) = \delta(x-1/2)$.

%----------%----------%----------%----------%----------
%\vspace{0.5cm}
\section{Numerical results and discussions}
\bs
Using the moments obtained above and performing the inverse Mellin
transformation (or trial and error),
we arrive at an expression of the quark momentum
distribution function $q(x)$.   Practically, it needs calculations
up to the
$4 \sim 6$-th moments to determine a shape of the structure
function by using eq. (\ref{Mellin}).
The calculated distribution function is defined at the low
energy scale, where the effective theory is supposed to work.
To compare with experiments, we
carry out the $Q^2$ evolution of the $n$-th moments by using the
perturbative QCD\cite{Muta}.
The result of the effective theory plays a boundary condition for
the QCD evolution.   Here, we choose a low energy scale
$\mu = 0.35 \GeV$,
which is a value of the constituent quark mass,
and the QCD scale parameter $\Lambda_{QCD} = 0.25 \GeV$.
It is difficult to estimate ambiguities of
the QCD evolution from such a low momentum scale.   It needs more
detailed studies to clarify use of the perturbative QCD below 1 GeV.

We show in Fig.3 the valence quark distribution function at $8
\GeV^2$\footnote{Of course, the valence antiquark distribution
has the same shape.}.  The
solid curve denotes the model calculation at zero density;
$\rho = 0$, which is to be compared with the
experimental fit indicated by the dashed curve\cite{SMRS}.
The result is in a reasonable agreement with the experiment.

We next show the results at the nuclear matter density
$\rho = \rho_0 $ in Fig.4 and Fig.5.
To deal with the medium effect, we use two different methods.

\vglue 0.4cm
SET1 : Simple application of the NJL model at finite density.

\in
SET2 : Brown-Rho scaling relation

\vglue 0.4cm
To calculate results of SET1, we simply use the NJL model at finite
density, described previously.
Density dependence of the various properties is calculated
by solving eqs.
(\ref{Gap}) and (\ref{T-pion}).  As the matter density increases,
the quark condensate, constituent quark
mass, and pion decay constant decrease about $30\%$,
whereas the pion mass and the
pion-quark-quark coupling constant are almost unchanged.
Such a behavior was already studied by several groups, and
details are found in ref. \cite{Kle}.

However, the confinement of quarks are absent in the NJL model, and so
this model at finite density implicitly assumes the `deconfined' quark
matter in the nucleus.
Hence, the result obtained by the NJL model at finite density is
not so reliable.
Here, we adopt another possible method to incorporate the density
dependence, namely, universal scaling law proposed by Brown and
Rho\cite{BR}.  Thus, we assume the following scaling relation;
\beq
\frac{M^*}{M} = \frac{g_{\pi qq}^*} {g_{\pi qq}} =
\frac{f_\pi^*}{f_\pi}
\;\; ,
\eeq
where the asterisk denotes the values of quantities in the medium.
In this case, the pion mass is assumed to be constant.
We use $M^* / M = 0.8$ at the normal nuclear matter density.

Results for the quark distribution function at the normal nuclear
matter density are shown in Fig.4.
Ratio of the quark distribution in the medium to one in the vacuum
is also shown in Fig.5.
As for SET1 indicated by
the solid curve, density
dependence of the quark distribution is weak.  The calculated
distribution function is almost same as one of the free pion shown in
Fig.3.
Only around $x \sim 1$, the quark distribution function decreases about $10
\%$.   Indeed, values of the 3-rd and 4-th moments change only
a few $\%$.
Such a weak density dependence is easily understood in the chiral
limit.
If we take the zero current mass $m=0$, the resulting pion mass
becomes
zero $m^2_\pi = p^2=0$ due to the Goldstone boson nature.
Inserting $p^2 = 0$ into
(\ref{gpqq}) and (\ref{I3}), functions $g_{\pi qq}$ and
$I_3$ are rewritten as,
\beq
g_{\pi qq}^{-2}(p^2=0 ) = -\frac{d}{dp^2} [p^2 I_2 (p^2)]
\to \frac{N_c}{4 \pi^2} \int ^{4( M^2 + \Lambda^2)}
_{4 (M^2 + {k_F}^2)} \frac{1}{ \kappa^2} \sqrt{1 - \frac{4
M^2}{\kappa^2}}    \;\;,
\eeq
\beq
I_3 (p^2=0) \to \frac{1}{6} \int ^{4( M^2 + \Lambda^2)}
_{4 (M^2 + {k_F}^2)} \frac{1}{ \kappa^2} \sqrt{1 - \frac{4
M^2}{\kappa^2}}   \;\; .
\eeq
Namely, $g_{\pi qq}^{-2}$ and $I_3$ are reduced to the same
function in the chiral
limit.
Using this result with eqs. (\ref{3rd_momA}) and (\ref{4th_momA}),
we find the 3-rd and 4-th moments become just numerical numbers,
\beq
A_3^{val} = \frac{1}{4} + \frac{1}{12} = \frac{1}{3}
\label{3rd_cl}
\eeq
\beq
A_4^{val} = \frac{1}{8} + \frac{1}{8} = \frac{1}{4} \;\; ,
\label{4th_cl}
\eeq
and thus density dependence of both moments vanishes.
Therefore, the valence quark distribution function in the chiral limit
(zero mass pion) is independent of the matter density.
In addition, from the results (\ref{1st_momA}, \ref{2nd_momA},
\ref{3rd_cl}, \ref{4th_cl}), we can write an analytical expression
of the quark distribution for the zero mass pion,
$ x q(x) = x \; (0<x<1)$ at the model scale,

In the case of SET2, the medium effect on the distribution function is
drastic as clearly seen in Fig.4 and 5.  For $x>0.6$, the result at the
nuclear matter density shows a substantial reduction.  In this case,
a value of the 3-rd moment decreases about $10 \%$, and $20 \%$
for the 4-th moment.

Finally, we discuss a connection between the nuclear pion enhancement
of the sea quark distribution  and the medium modification of the
pion structure function.
{}From eq. (\ref{pion_nucl}) calculated by the random phase
approximation, it can be shown that the pion momentum
distribution in the nucleus $f_{\pi NN}^A (y)$ is peaked at small $y$;
$y<0.3$\cite{ET,pion_excess}.
Thus, the integral of eq. (\ref{convl}) is dominated by the small $y$
contribution.  Let us recall that argument of
$\bar q_\pi (x_\pi =x/y)$ is proportional to $1/y$.
Hence, the integral (\ref{convl}) is sensitive to the large $x_\pi$
behavior of the quark distribution function\cite{NPE}.
We have shown that the quark distribution function at the large $x_\pi$
decreases as the density increases.  Hence, we expect that the reduction
of the pion structure function at the large Bjorken-$x$ in the medium
compensates the enhancement of the sea quark distribution due to
the pion excess in the nucleus.
In fact, it can be shown that, for $x>0.1$,
the enhancement of the sea in the nucleus almost
disappears by using the density dependent pion structure function
of SET2.  On the other hand, the nuclear pion
enhancement of the sea quark still remains in the case of SET1.  More
detailed discussions will be given in the forthcoming paper\cite{NPE}.

In conclusion, we have studied the pion structure function in the
nuclear
medium within the NJL model as an effective theory of QCD.  We have
calculated the moments of the pion structure function based on OPE to
separate the soft part of the matrix elements and to
maintain the gauge invariance.
The result for the free pion shows a reasonable agreement with the
experiment.
We have shown that the quark distribution at the large $x$ region
decreases in the nuclear medium $\rho = \rho_0$.
Magnitude of the modification depends on treatments of the medium
effect.  We have found that the medium modification of the
pion structure
function is small within the NJL model around the nuclear
matter density.
Instead, if we use the Brown-Rho scaling relation,
the quark distribution
function shows  substantial reduction for the large $x$.
We note that the quark distribution function in the high density limit
approaches the non-relativistic one, $q (x) = \delta (x-1/2)$ (at the
model scale).
It is also interesting to study the medium modification of other meson
structure functions, which is under consideration\cite{NPE}.

We have also discussed the nuclear pion enhancement
of the sea quarks in the nucleus in terms of the density dependent pion
structure function.
The reduction of the pion structure function at the large $x$ may
lead a
suppression of the sea quark distribution due to the pion cloud.
Such a tendency seems to be consistent with the available
experiments, where the sea quark enhancement in the nucleus is not
observed.

%%%%%%%%%%%%%%%%%%%%% REFERENCES %%%%%%%%%%%%%%%%%%%%%%%%%%

%
%
\new
\ni
{\large Figure Captions}

\vspace{0.5cm}
\ni
Fig. 1

\ni
The nucleon-hole and delta-hole excitations contribute to the
deep inelastic lepton nucleus scattering.  Solid line denotes the
nucleon, and the double one the delta.  The pion and virtual photon are
depicted by the dashed and wavy curves, respectively.

\vspace{1cm}

\ni
Fig. 2

\ni
Graphical representation of the operator product expansion.
The quark and the pion are depicted by the solid and dashed
curves, respectively.  Wavy curve denotes the virtual photon.
Left hand
diagram shows the forward scattering amplitude of the pion in the NJL
model.

\vspace{1cm}

\ni
Fig. 3

\ni
The valence quark momentum distribution function at $Q^2 = 8
\GeV^2$ as a function of the Bjorken $x$.
The solid curve denotes theoretical calculation for the free
pion ($\rho = 0$), and the dashed curve the experimental
fit\cite{SMRS}.

\vspace{1cm}

\ni
Fig. 4

\ni
The valence quark distribution function of the pion in the nuclear
medium $\rho=\rho_0$ at
$Q^2 = 8 \GeV^2$.  The result of SET1 is depicted by the solid curve,
and SET2 by the dashed curve.   See text for detail.

\vspace{1cm}

\ni
Fig. 5

\ni
Ratio of the quark distribution functions at the normal nuclear matter
density ($\rho= \rho_0$)
to one of the free pion ($\rho = 0$);
$q(x)_{\rho=\rho_0} / q(x)_{\rho=0}$.
Results of SET1 and SET2 are depicted by solid and dashed curves,
respectively.

\end{document}